\documentclass[runningheads]{llncs}
\usepackage[T1]{fontenc}
\usepackage{graphicx}
\usepackage{color}
\usepackage{booktabs}
\usepackage{multirow}
\begin{document}
\title{Zero-Shot vs. Few-Shot Multi-Speaker TTS Using Pre-trained Czech SpeechT5 Model}
\titlerunning{Multi-Speaker TTS Using SpeechT5 Model}
%
\author{Jan Lehečka\inst{1}\orcidID{0000-0002-3889-8069} \and
Zdeněk Hanzlíček\inst{1}\orcidID{0000-0002-4001-9289} \and
Jindřich Matoušek\inst{1,2}\orcidID{0000-0002-7408-7730} \and
Daniel Tihelka\inst{1}\orcidID{0000-0002-3149-2330}
}
\authorrunning{J. Lehečka et al.}
%
\institute{NTIS – New Technologies for the Information Society, Faculty of Applied Sciences, Pilsen, Czech Republic,
\email{\{jlehecka,zhanzlic,jmatouse,dtihelka\}@ntis.zcu.cz} \and
Department of Cybernetics, Faculty of Applied Sciences, University of West Bohemia in Pilsen, Czech Republic,
\email{jmatouse@kky.zcu.cz}}

\maketitle              
\begin{abstract}
In this paper, we experimented with the SpeechT5 model pre-trained on large-scale datasets. We pre-trained the foundation model from scratch and fine-tuned it on a large-scale robust multi-speaker text-to-speech (TTS) task. We tested the model capabilities in a zero- and few-shot scenario. Based on two listening tests, we evaluated the synthetic audio quality and the similarity of how synthetic voices resemble real voices. Our results showed that the SpeechT5 model can generate a synthetic voice for any speaker using only one minute of the target speaker's data. We successfully demonstrated the high quality and similarity of our synthetic voices on publicly known Czech politicians and celebrities.

\end{abstract}

\section{Introduction}

Text-to-speech (TTS) systems have improved significantly in recent years thanks to the rapid development of deep learning methods, end-to-end modeling, and the availability of large training datasets. Motivated by the recent success of pre-trained models for other modalities (e.g., large language models for text or Wav2Vec models for speech \cite{baevski2020wav2vec}), the natural trend in solving the TTS task is to pre-train a multi-speaker TTS model from a large extensive speech dataset and then to exploit this pre-trained knowledge to obtain a TTS system with the desired voice and style using only a few examples of the target voice \cite{casanova2022yourtts,wang2023neural}.
This is in contrast to the common single-speaker TTS approaches, in which usually a large and high-quality clean dataset must be recorded for each speaker in a recording studio to train a TTS model from scratch.

The approach of using only a small amount of the target speaker's data (usually a few seconds or minutes of speech) to fine-tune the models is called a \textit{few-shot multi-speaker TTS}. An extreme case when the system can synthesize speech in the voice of the target speaker without any additional fine-tuning is called \textit{zero-shot multi-speaker TTS}. 
In this paper, we are investigating both approaches using the pre-trained Czech SpeechT5 model \cite{ao-etal-2022-speecht5}, a multi-modal model pre-trained from a large amount of unlabeled text and speech. Other popular large speech language models enabling a high-quality zero-/few-shot synthesis are YourTTS \cite{casanova2022yourtts}, TorToiSe \cite{Betker_TorToiSe_text-to-speech_2022}, Vall-E \cite{wang2023neural}, StyleTTS2 \cite{styletts2}, Voicebox \cite{le2023voicebox}, Mega-TTS2 \cite{jiang2024megatts}, UniAudio \cite{yang2023uniaudio}, HierSpeech++ \cite{lee2023hierspeech}, or NaturalSpeech2/3 \cite{shen2023naturalspeech,ju2024naturalspeech}. 

The SpeechT5 model is suitable for both zero- and few-shot TTS tasks. The zero-shot TTS is achieved by inputting the speaker embedding along with the text to be synthesized. The input embedding could be a vector derived from a real human speech or a completely artificial vector (e.g., a random vector or interpolation between different speaker embeddings). A few-shot TTS is typically used when the target voice or style is somehow specific or too far from the data observed during the model training. In such cases, several speech examples must be collected for short additional fine-tuning of the model.

Zero- and few-shot multi-speaker TTS enables and accelerates the production of new synthetic voices, even of speakers not observed in the training data, which is highly desirable in many applications. To name some of them, few-shot TTS models could preserve the voice of laryngectomy patients \cite{tihelka2021}, zero-shot models could generate new production voices without copyrighting, both zero- and few-shot models could enrich conversational and dialogue systems with a lot of new voices, and using voice-cloning systems, users and developers can easily generate technology that speaks in their own voice.

In this paper, we are experimenting with TTS systems that produce the voices of publicly known Czech politicians and celebrities. Experimenting with these voices has several advantages: (1) their voices are publicly well-known, so our evaluators could compare how synthetic voices resemble real voices more confidently, (2) a satisfactory amount of fine-tuning data could be automatically collected from their public speeches or interviews, (3) we can experiment with voices that belong -- in contrast to common TTS systems -- to non-professional speakers, even with voices that are very original and specific.

On the other hand, such synthetic voices of famous celebrities and politicians pose a great risk of being misused, for example, to spread disinformation or fake recordings. For this reason, we decided not to release our final TTS models to the public. 
However, to allow the speech community to experiment with our foundation model and verify our findings, we are releasing the pre-trained Czech SpeechT5 model for non-commercial use\footnote{\url{https://huggingface.co/fav-kky/SpeechT5-base-cs-tts}}.

\section{SpeechT5}
The development of the SpeechT5 model \cite{ao-etal-2022-speecht5} was motivated by the success of the pre-trained language model T5 (Text-To-Text Transfer Transformer) \cite{2020t5}. SpeechT5 is a multi-modal extension of the (text-only) T5 model. It is pre-trained jointly on two modalities -- text and speech.
The SpeechT5 framework adopted the encoder-decoder Transformer architecture \cite{NIPS2017_3f5ee243} with additional six
modal-specific (speech/text) pre/post-nets. 
The speech/text pre-nets preprocess the input into latent representations, which are fed into a shared encoder-decoder network to apply the sequence-to-sequence transformation. Finally, the post-nets generate the output in the speech/text modality based on the output of the decoder. The speech modality is trained on the raw audio signals on the model's input and  80-dimensional log Mel-filterbank representations on the output, which is followed by a pre-trained HiFi-GAN vocoder to get the final output audio signal.

The SpeechT5 model is pre-trained from large-scale unlabeled speech and text data using self-supervised learning. 
During the pre-training, a cross-modal vector quantization approach is used to mix up speech/text states with latent units randomly. This approach aligns the textual and speech information into the unified semantic space.

After pre-training the SpeechT5 model, a whole range of possibilities opens. It can be fine-tuned on a very large pallet of text- and speech-related tasks, including TTS, automatic speech recognition (ASR), speech translation (ST), question answering (QA), speech identification (SID), voice conversion (VC), speech denoising (SD), speech enhancement (SE), chatbots and many others.

To match individual speakers with audio signals, speaker embeddings are used in the SpeechT5 model. During both pre-training and fine-tuning on the TTS task, a speaker embedding derived from the target signal is also input along with each audio training example. This allows the model to use information encoded in the speaker embedding vectors when generating new signals. Specifically, x-vectors \cite{snyder2018spoken} are used to encode individual speakers in this model.

\subsection{Pre-training}
The only public SpeechT5 model we are aware of is for the English language only \cite{ao-etal-2022-speecht5}. Transferring the knowledge from the English model into Czech tasks via fine-tuning did not yield satisfactory results. Since we were able to collect a sufficient amount of Czech unlabeled data and since we have access to a high-end GPU cluster, we decided to pre-train our own monolingual model from scratch and release it to the public.

Self-supervised learning depends on a large amount of unlabeled training data, and transformers are known to scale their performance well with the size of pretraining data, even with huge datasets \cite{2020t5,babu2021xls}. Hence, we tried to gather as much public and in-house unlabeled data as possible from both target modalities (speech and text). 

\subsubsection{Speech Data} We collected more than 120 thousand hours of Czech speech. We are not aware of any similar collection of Czech speech data at this scale mentioned in the literature so far. The collection includes recordings from TV shows (31k hours), radio broadcasts (27k hours), podcasts (24k hours), VoxPopuli dataset \cite{wang-etal-2021-voxpopuli} (18.7k hours), shadow speakers (12k hours), sports (5k hours), telephone data (4k hours), and a smaller amount of data from several other domains. Although the majority of this collection is from public sources, we don't have the license to release it publicly.

Since the feature extraction of the input signal is limited by the memory of GPUs in use, we sliced all records not to exceed 30\,s, which we found to be the maximum input length we were able to fit in the GPU memory.

\subsubsection{Text Data} As a source of unlabeled text data, we used the Common Crawl project\footnote{\url{https://commoncrawl.org}} which is a huge public web archive consisting of petabytes of crawled web pages.
We used the language information provided in the index files to select Czech records only. The corresponding plain texts (stored in WET archives) were downloaded. To clean the data, we followed almost the same rules that were used to pre-process the English C4 dataset for pre-training the T5 model \cite{2020t5}, i.e.:
\begin{enumerate}
    \item We only retained lines that ended with a terminal punctuation mark (“.”, “?” or “!”).
    \item We removed lines containing “javascript” or “cookies”.
    \item As a rough filter for offensive content, we removed web pages containing any offensive word from a black-list\footnote{\url{https://github.com/LDNOOBW/List-of-Dirty-Naughty-Obscene-and-Otherwise-Bad-Words}}.
    \item We removed web pages containing strings “lorem ipsum” or “\{”.
    \item We retained only pages classified as Czech with the probability of at least 0.99 according to \texttt{langdetect}\footnote{\url{https://pypi.org/project/langdetect/}} tool.
    \item To deduplicate the dataset, we discarded all but one line occurring more than once in the data set.
    \item We only retained lines with at least 3 words and pages with at least 5 sentences.
\end{enumerate}
This simple yet rigorous cleaning process removed about 98\% of downloaded plain texts from the dataset, mainly due to the deduplication (the more data we downloaded, the harder it was to find a new unobserved line).
We downloaded and cleaned crawls from August 2018 to October 2021. Together, we processed 35 crawls, and the resulting dataset contains text from 530 million web pages with 17.5 billion words (125~GB of cleaned text). Finally, we converted all texts into lowercase.

\subsubsection{Pre-training Setup} We used the same pre-training setup and hyperparameter values as for the base SpeechT5 model in \cite{ao-etal-2022-speecht5}. We pre-trained the model for 500 thousand steps with a batch size of about 1.6 hours of audio and 770 thousand text characters (approx. 100 thousand words). Ultimately, the model iterated about $6\times$ over the full speech dataset and $3\times$ over the full text dataset during the whole pre-training. 
To pre-train the model, we used the implementation released along with the original SpeechT5 paper\footnote{\url{https://github.com/microsoft/SpeechT5}} and prepared the data in the same way. The pretraining took about 18 days on a machine with eight NVIDIA A100 GPUs.

\subsection{Fine-tuning}
Since the pre-trained SpeechT5 model is trained only with self-supervised learning to predict missing pieces of input data, we needed to fine-tune the model to multi-speaker TTS task first. We aimed for a large, clean, and diverse multi-speaker fine-tuning dataset containing various voices and speech styles across many different speech domains (spontaneous speech, read speech, narration, oration, etc.). We used a combination of several ASR datasets, one TTS dataset, and a large amount of automatically transcribed speech data from public radio broadcasts. Specifically, we used the following data sources:

\begin{itemize}
    \item \textbf{Public ASR datasets} -- We used annotated speech data from the Czech portion of the CommonVoice dataset (denoted as \textit{ASR-CV}) in version 7.0 containing 49 hours of validated speech \cite{commonvoice:2020}, VoxPopuli dataset (\textit{ASR-VP}, 62 hours) \cite{wang-etal-2021-voxpopuli}, and Czech Speecon database (\textit{ASR-Speecon}, 733 hours)\footnote{\url{https://catalogue.elra.info/en-us/repository/browse/ELRA-S0298/}}. The CommonVoice is a crowd-sourced dataset collected by Mozilla containing mostly read sentences from almost 500 speakers. In contrast, the VoxPopuli dataset is a speech corpus collected from 2009-2020 European Parliament event recordings, so it contains mostly politicians' public speeches (or their simultaneous translations) uttered by 138 different speakers. Finally, the Czech Speecon database contains phonetically rich sentences uttered by a balanced mix of 600 speakers (including also children's voices) recorded on several microphones with different levels of background noise.
    \item \textbf{In-house ASR datasets} -- \textit{SPT-MGW} is a large-scale in-house collection of non-professional voices recorded in a common environment on a common microphone for robust ASR training purposes. It contains 382 hours of read speech from 1,115 different speakers.
    \item \textbf{In-house TTS datasets} -- \textit{TTS-PRO} is a high-quality in-house phonetically rich TTS dataset recorded by one professional speaker in a recording studio. It contains 16 hours of a very clean speech.
    \item \textbf{Public Radio Broadcasts} -- To increase the diversity of speakers and speech styles, we downloaded a lot of audio data from public radio broadcasts. We used a voice activity detector (VAD) to automatically select only parts containing speech and slice long signals on pauses into short segments not exceeding 30 seconds. Then, we used the in-house ASR system specifically tailored to the radio broadcast domain to get the transcription of each segment. We kept only segments with high ASR confidence for all generated words in the transcripts. In other words, we discarded all segments where the in-domain ASR model was not sure about the correct transcription. This way, we collected almost 4,000 hours of transcribed speech from this source. 
\end{itemize}

\subsubsection{Data Preprocessing}
\label{sec:ft_preproc}
Almost all mentioned datasets were not entirely suitable for the TTS task and needed some pruning. For example, ASR datasets contain noisy and hard-to-understand examples, which are important and challenging for the ASR task but undesired for the TTS task (we want the generated speech to be as clean as possible). Therefore, we designed a pipeline of preprocessing filters and algorithms to select only clean and correctly transcribed examples while discarding all noisy, erroneous, or otherwise problematic examples.

\begin{enumerate}
    \item First, we noticed that many records contained long pauses at the beginning or end of the signal, which confuses the model as there is no reason why some sentence starts immediately in the record and some after a long pause.
    To make sure our training records do not contain too long pauses at the beginning or end, we used the Wav2Vec-based ASR system to predict the timestamps of the first and the last word and trimmed the leading and trailing pauses not to exceed 0.25 seconds. We experimented also with padding too short pauses with silence, but it did not result in better performance.
    \item To filter out examples with too high background noise, we adopted the same cleaning criterion as used for the clean LibriTTS dataset \cite{zen2019libritts} and using WADA-SNR algorithm \cite{kim08e_interspeech}, we discarded all examples with speech-to-noise ratio lower than 20dB. 
    \item To filter out transcription errors, we validated each transcription against the output from the state-of-the-art general-purpose ASR model \cite{wav2vec2-base-cs-80k-ClTRUS} and discarded all segments where the character error rate (CER) exceeded 0.1.
    \item Some transcripts did not contain punctuation, which could be an important clue for the TTS system to generate pauses and breaths correctly. We used the T5 model fine-tuned on the punctuation restoration task \cite{t5punc} to restore punctuation in the transcripts. After that, we added a full stop at the end of each transcript, for which the T5 model failed to determine a correct terminal punctuation mark (“.”, “?” or “!”). 
    Ensuring each transcript ends with a terminal punctuation mark enables the TTS model to use it as a clue to learn to stop the speech generation process at the correct time.
    \item We cleaned the transcripts by removing non-speech events and converting all words into lowercase to keep the same format as in the pre-training data.
    \item We discarded too-short examples (shorter than 1s), too-long examples (longer than 30s), and all examples with empty transcripts.
    \item Finally, we generated a speaker embedding vector for each remaining example using an x-vector model released by SpeechBrain\footnote{\url{https://huggingface.co/speechbrain/spkrec-xvect-voxceleb}}.
\end{enumerate}

After the preprocessing, we ended up with a large-scale diverse multi-speaker TTS dataset with over a million clean audio recordings with a total duration of 1,668 hours. We tabulate the statistics about this dataset that was used for fine-tuning the SpeechT5 model in Tab. \ref{tab:tab_ft_data}.

\begin{table}
\caption{Statistics of the multi-speaker fine-tuning TTS dataset after preprocessing. We show the number of speech hours, the number of audio files, and the number of words in the transcripts.}
\label{tab:tab_ft_data}
\centering

\begin{tabular}{l l r r r r r r r}
\toprule
\multirow{2}{*}{dataset} & \multirow{2}{*}{domain} & \multicolumn{3}{c}{train data} & & \multicolumn{3}{c}{validation data} \\
\cline{3-5} \cline{7-9}
 &  & hours & files & words & & hours & files & words \\
 \midrule
ASR-CV & read sentences & 21.6 & 25,180 & 173,944 & & 5.0 & 5,603 & 39,787 \\
ASR-VP & public speeches & 39.0 & 13,753 & 299,897 & & 2.4 & 863 & 18,504 \\
ASR-Speecon & balanced mix & 92.9 & 151,531 & 548,968 & & 1.9 & 3,131 & 11,070 \\
ASR-SPT-MGW & read sentences & 145.2 & 115,121 & 1,039,934 & & 1.5 & 1,188 & 10,445 \\
TTS-PRO & professional & 15.7 & 12,148 & 119,119 & & 0.0 & 20 & 153 \\
Radio & radio broadcasts & 1,353.9 & 711,897 & 11,025,299 & & 1.2 & 698 & 10,139 \\
 \midrule
TOTAL &  & 1,668.3 & 1,029,630 & 13,207,161 & & 12.0 & 11,503 & 90,098 \\
\bottomrule
\end{tabular}
\end{table}

\subsubsection{Fine-tuning Setup}
To fine-tune the SpeechT5 model, we used Transformers library\footnote{\url{https://huggingface.co/docs/transformers}}. We trained with a batch size of 256 examples for 120 thousand steps. The learning rate was warmed up for the first 10,000 steps to a maximum value of $1\times10^{-4}$ and then decayed linearly to zero for the rest. We left all other fine-tuning hyperparameters to the default values. The fine-tuning took 31 hours on one NVIDIA H100 GPU, and the model iterated almost $28\times$ over the whole dataset.

\section{Few-Shot Speech Data Collection}
To test our TTS system on diverse voices, we selected 15 real human speakers.
For each selected speaker, we collected a small amount of speech data. We aimed for the best audio and speech quality possible, so we searched mainly for speeches recorded in silent rooms on high-quality microphones.
Based on the source of the selected speech data for few-shot fine-tuning, we distinguish between 3 types of speech data:
\begin{itemize}
    \item \textbf{Oration} -- A major public speech addressed to the whole nation on the occasion of some important event, such as the President's New Year's speech. These speeches are typically not spontaneous but are read from a reading device; however, the speakers usually aim for an emotional and solemn speech. A typical duration of used orations is from five to ten minutes.
    \item \textbf{Interview} -- An interview of the target speaker with a reporter. We searched for interviews with low background noise and a duration of about 30 minutes to ensure there would be at least 5 minutes of clean speech from the target speaker. We used mainly public interviews broadcast on the radio. Speech from interviews is spontaneous and can contain disfluencies, unfinished sentences, imperfect pronunciation, and non-speech events such as laughter, coughing, etc.
    \item \textbf{Read speech} -- A collection of spoken sentences recorded in a recording studio. Voices in this group belonged to publicly unknown non-professional speakers, whose data was self-recorded at our department.
    We included this group of speakers because it is certain that they were not part of the training data, not a tiny bit. This is not certain for the other groups of speakers, as one of the data sources was public radio broadcasting containing the voices of a rich mix of famous public figures. However, we made sure the test sentences used in the final listening tests were not part of any training dataset used to train the model.
    
\end{itemize}

Together, we collected data for 5 voices of the \textit{read-speech} type (4 males, 1 female), 4 voices of the \textit{oration} type (4 publicly well-known male voices of politicians), and 6 voices of the \textit{interview} type (5 female voices and 1 male, all publicly well-known figures and celebrities).
To preprocess the collected data of the \textit{read-speech} type, we followed the same rules as for the large-scale fine-tuning dataset (see Sec. \ref{sec:ft_preproc}). Orations and interviews had to be preprocessed differently since we didn't have annotations or transcripts for them. 

For \textit{oration} and \textit{interview} data type, we first segmented the speech into short segments. This was achieved by transcribing the speech using the ASR model \cite{wav2vec2-base-cs-80k-ClTRUS} followed by punctuation reconstruction \cite{t5punc}. We then scored all pauses between words based on their duration and the presence of terminal punctuation marks and found optimal segmentation into short segments (shorter than 30 seconds). This way, we obtain short speech segments with transcript hypotheses, which was then preprocessed the same way as the \textit{read-speech} type.

In the case of \textit{interviews}, we had to apply an additional filter to remove the reporter's voice from the data. At the end of the preprocessing, we checked each record for speaker verification with the target voice. To do this, we had to find one representative example of the target speaker's voice and compare each segment with this reference voice. We kept only sentences belonging to the target speaker. We used the ECAPA-TDNN model released by SpeechBrain\footnote{\url{https://huggingface.co/speechbrain/spkrec-ecapa-voxceleb}} for the speaker verification.
Finding a short representative example of interview voices was the only manual work we had to do; otherwise, the whole pipeline of data collection and preprocessing for all three types of voices was fully automatic.

\section{Results and Discussion}

For each selected voice, we held two short sentences out of the training datasets to ensure the model had not observed the records during the training. After fine-tuning the model for all tested scenarios and target voices, we generated the same sentences using the SpeechT5 model. When generating each sentence, the model expected a speaker embedding as additional input. We randomly selected one embedding from the speaker's fine-tuning data. 
We measured the quality of synthetic voices and the speaker similarity (i.e., how synthetic voices resemble real voices) by performing listening tests. 
For a better idea about the quality and similarity of the evaluated recordings, we published some recordings for comparison and listening along with the pre-trained SpeechT5 model\footnote{\url{https://huggingface.co/fav-kky/SpeechT5-base-cs-tts}}.

\subsection{Listening tests}

Despite the significant progress in the automatic evaluation of speech quality in recent years~\cite{cooper_2023_asru}, listening tests with human listeners remain the fundamental method for assessing various complex speech properties, such as quality or voice identity. Therefore, we organized two independent listening tests; each focused on one of the aforementioned characteristics, i.e., quality and identity/similarity of generated speech.  Both tests were created using a web-based framework~\cite{gruber_2019_interspeech}. Ten listeners took part in both tests; they could replay each utterance multiple times and then simply set the perceived quality/similarity using vertical sliders.

We selected 15 various voices (4$\times$oration, 6$\times$interview, and 5$\times$read speech).
Both tests contained 2 sets for each voice, i.e., 30 sets of utterances together. Each set comprised a reference (natural) recording and 4 generated utterances for the evaluation; we included zero-shot and fine-tuning with 10 seconds, 1 minute, and 5 minutes of speech data.

In the quality listening test, the reference recordings were also added for evaluation. However, we did not construe the reference recordings as a top-quality goal~\footnote{Reference recordings are often considered as the highest achievable quality in the MUSHRA~\cite{mushra_2014} listening test.} that should be achieved since some source voices contained various acoustic or pronunciation imperfections; therefore, listeners could prefer generated utterances to natural speech. In the similarity listening test, the participants should only judge the similarity with the reference recording; the quality should not be considered.

\begin{figure}[t]
\center
\includegraphics[width=1.0\textwidth]{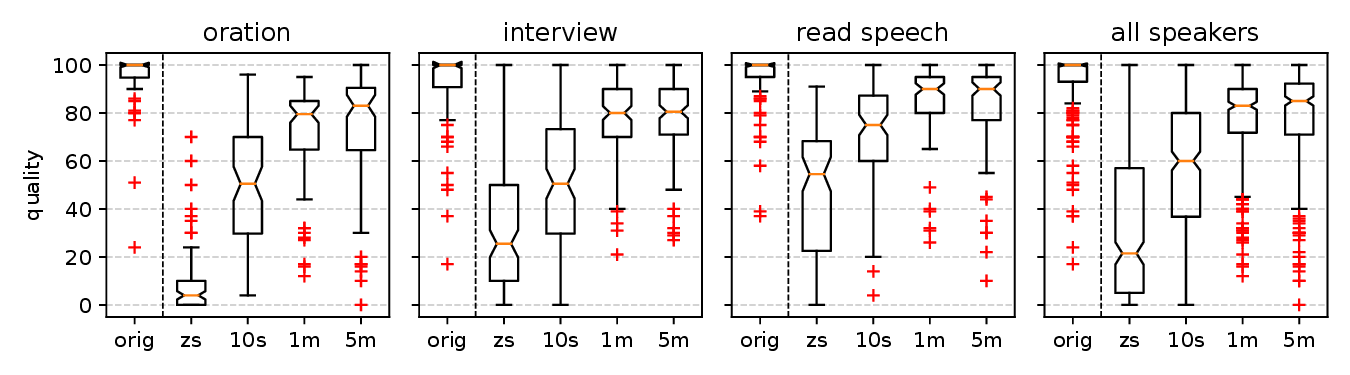}
\caption{Results of the quality listening test. We show the results for the original real-human records (\textit{orig}) and for the records generated using the SpeechT5 model: the zero-shot approach (\textit{zs}) and few-shot approaches using 10 seconds of training data (\textit{10s}), one minute (\textit{1m}), and five minutes (\textit{5m}) respectively.}
\label{fig:test_quality}
\end{figure}

\begin{figure}[t]
\center
\includegraphics[width=1.0\textwidth]{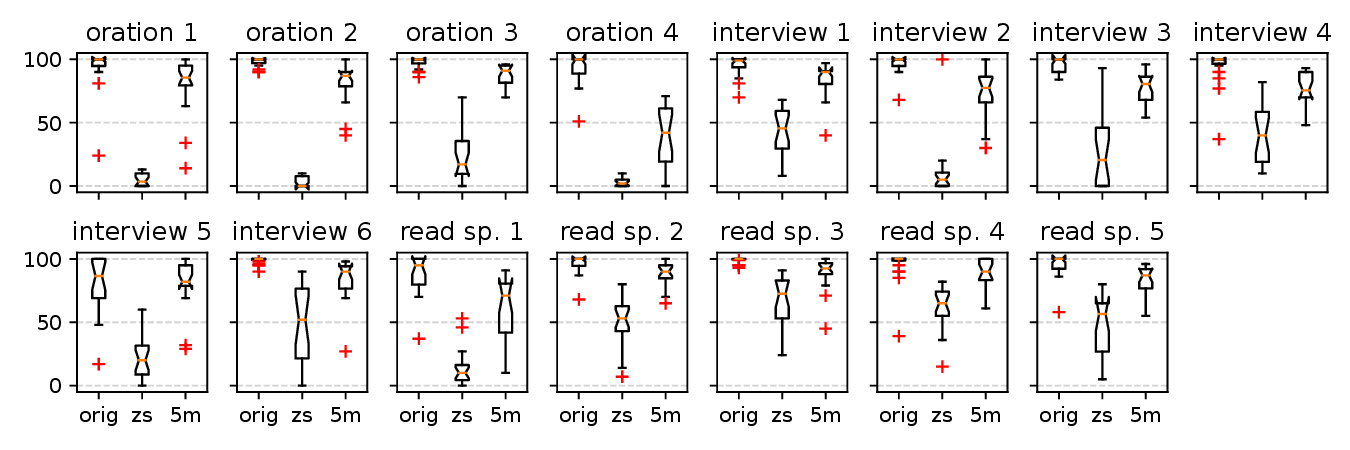}
\caption{Quality listening test -- results for individual voices.}
\label{fig:test_quality_individual}
\end{figure}

The summary results of the quality listening test are presented in Fig.~\ref{fig:test_quality}.
We can see that the zero-shot approach (abbreviated as \textit{zs} in our figures) works poorly with the SpeechT5 model. The fine-tuned models perform significantly better. Approximately 1 minute of data seems sufficient to achieve a good quality of generated speech. Further increasing the amount of data has generally only a minor impact. Simplified evaluation for individual voices is presented in Fig.~\ref{fig:test_quality_individual}. These results have no statistical significance, but the graphs illustrate the variability of experimental voices, e.g., an uneven source data quality, which is a reason for many distant outliers in summary graphs.

\begin{figure}
\center
\includegraphics[width=1.0\textwidth]{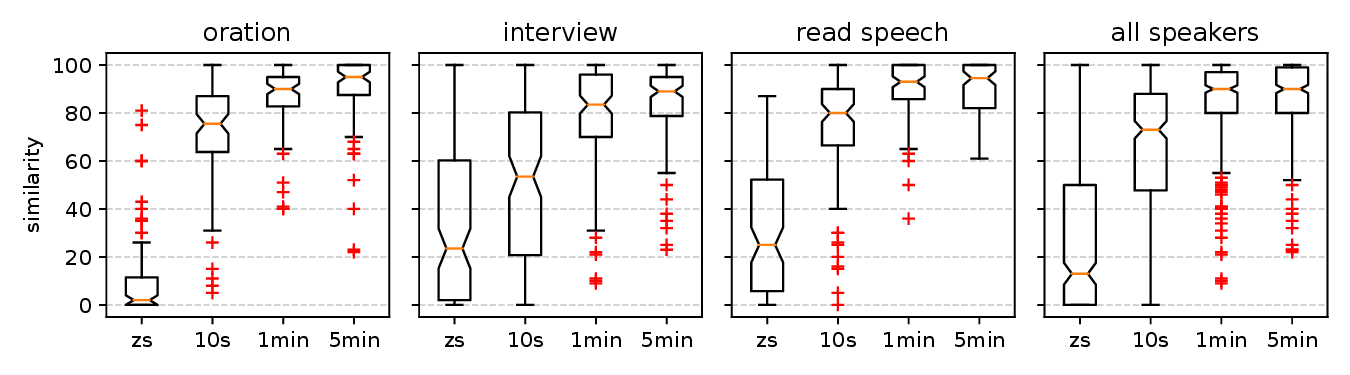}
\caption{Results of the similarity listening test.}
\label{fig:test_similarity}
\end{figure}

The results of the similarity listening test are presented in Fig.~\ref{fig:test_similarity}. In agreement with the first test, zero-shot did not perform well. However, the similarity continued increasing with more fine-tuning data for less monotonous voices (``pathetic'' orations and ``spontaneous'' interviews).

\section{Conclusion}
In this paper, we experimented with the SpeechT5 model pre-trained on large-scale datasets. We selected voices of publicly known Czech politicians and celebrities and tested the model in a zero- and few-shot multi-speaker TTS task.
Our listening tests results suggest that the zero-shot performance of the fine-tuned SpeechT5 TTS models was very poor, both in the quality and similarity aspects. On the other hand, the few-shot performance of the models seems like a promising way to go, as just after a short fine-tuning from one minute of the target speaker's speech, the quality and similarity seemed acceptable for most listeners. Adding more speech data into fine-tuning did not further improve the quality but 
slightly further improved the similarity of the target voice, especially for less monotonous voices.

\begin{credits}
\subsubsection{\ackname}
This research was supported by the Czech Science Foundation (GA CR), project No. GA22-27800S. Computational resources were provided by the e-INFRA CZ project (ID:90254), supported by the Ministry of Education, Youth and Sports of the Czech Republic.

\subsubsection{\discintname}
The authors have no competing interests to declare that are relevant to the content of this article.
\end{credits}
%
%
%
\bibliographystyle{splncs04}
\bibliography{tsd1213a}
\end{document}